\documentclass[pre, twocolumn, superscriptaddress]{revtex4-1}
\usepackage{amsmath,amsbsy, amssymb}
\usepackage[dvips]{graphicx}
\usepackage{xcolor}
\usepackage{units}

\DeclareGraphicsExtensions{.eps,.ps}

\usepackage{bm}% bold math

\def\half{{\textstyle \frac{1}{2}}}

\def\Tr#1{{\textrm{ Tr}} \left( #1 \right)}
\def\Det#1{\textrm{ Det}\left( #1 \right)}
\def\det#1{\textrm{ Det}( #1 )} % with lc det one has small brackets

\def\ber{\begin{eqnarray}}
\def\eer{\end{eqnarray}}
\def\be{\begin{equation}}
\def\ee{\end{equation}}
\def\beno{\begin{equation*}}
\def\eeno{\end{equation*}}
\def\bea{\begin{eqnarray}}
\def\eea{\end{eqnarray}}

\begin{document}
%\draft \twocolumn[\hsize\textwidth\columnwidth\hsize\csname
%@twocolumnfalse\endcsname

\title{Finite wavelength surface-tension driven instabilities in soft solids, including instability in a cylindrical channel through an elastic solid.}

\author{Chen Xuan}
\affiliation{Department of Mechanics and Engineering Science, Fudan University, Shanghai 200433, China}
\affiliation{Cavendish Laboratory, Cambridge University, 19 JJ Thomson Avenue, Cambridge, United Kingdom}
\author{John Biggins}
\affiliation{Cavendish Laboratory, Cambridge University, 19 JJ Thomson Avenue, Cambridge, United Kingdom}
\date{\today}
%%%%%%%%%%%%%%%%%%%%%%%%%%%%%%%%%%%%%%%%%%%%%%%%%%%%%%%%%%%%%%%
\begin{abstract}
We deploy linear stability analysis to find the threshold wavelength ($\lambda$) and surface tension ($\gamma$) of Rayleigh-Plateau type ``peristaltic'' instabilities in incompressible neo-Hookean solids in a range of cylindrical geometries with radius $R_0$. First we consider a solid cylinder, and recover the well-known, infinite wavelength instability for $\gamma\ge6 \mu R_0$, where $\mu$ is the solid's shear modulus. Second, we consider a volume-conserving  (e.g.\ fluid filled and sealed) cylindrical cavity through an infinite solid, and demonstrate infinite wavelength instability for $\gamma\ge 2 \mu R_0$. Third, we consider a solid cylinder embedded in a different infinite solid, and find a finite wavelength instability with $\lambda\propto R_0$, at surface tension $\gamma \propto \mu R_0$, where the constants depend on the two solids' modulus ratio. Finally, we consider an empty cylindrical channel (or filled with expellable fluid) through an infinite solid, and find an instability with finite wavelength, $\lambda \approx2 R_0$, for  $\gamma\ge 2.543... \mu R_0$. Using finite-strain numerics, we show such a channel jumps at instability to a highly peristaltic state, likely precipitating it's blockage or failure. We argue that finite wavelengths are generic for elasto-capillary instabilities, with the simple cylinder's infinite wavelength being the exception rather than the rule.
\end{abstract}
%%%%%%%%%%%%%%%%%%%%%%%%%%%%%%%%%%%%%%%%%%%%%%%%%%%%%%%%%%%%%%%
%\vspace{0.2cm}
\pacs{46.25.Cc, 46.70.De, 46.90.+s, 83.80.Va}
% \narrowtext
 \maketitle
 %\begin{multicols}{2}
%%%%%%%%%%%%%%%%%%%%%%%%%%%%%%%%%%%%%%%%%%%%%%%%%%%%%%%%%%

%In this study, we investigate how a soft cylindrical solid deforms driven by surface tension. We will study both a cylinder with radius $R_0$ and a hollow cylinder with inner radius $R_0$ and outer radius which is much larger than $R_0$. The solid is un-deformed in the initial stress free state and we shall see how the presence of surface tension drives surface instability to occur. The instability is the result of the competition between the elastic energy and surface energy. The elastic energy penalizes deformation whereas the surface energy abhors big surface areas. The minimum surface area at a constant volume should be a sphere rather than the initial un-deformed state of the solid.
\section{Introduction}\label{sect:intro}
We are used to surface tension influencing the shape of fluids: it makes droplets round, fashions menisci and breaks apart fluid columns via the Rayleigh-Plateau instability \cite{strutt1879instability}. Conventionally we do not think of surface tension as sculpting solids, and with good reason: a solid's surface tension, $\gamma$, can only compete against its elastic shear modulus, $\mu$, at length-scales comparable to or below the solid's elastocapillary length-scale, $l_{cap}=\gamma/\mu$, which, for most crystalline solids, is sub-angstrom. However, in suitably soft solids, such as gels and biological tissues, $l_{cap}$ can be microns or even millimeters, making surface tension-driven distortions important in many biological contexts  and readily accessible in laboratories \cite{roman2010elasto}. Recent work has highlighted ``peristaltic'' surface-tension instabilities in soft solid cylinders \cite{matsuo1992patterns, barriere1996peristaltic, mora2010capillarity,ciarletta2012peristaltic,ciarlettanonlinearjmps, ciarlettanonlinearpre}, elastocapillary modifications to the theory of wetting \cite{style2013universal, style2013surface,style2013patterning}, capillarity-driven bending of wet elastic rods and sheets \cite{bico2004adhesion, kim2006capillary, mora2013solid}, and the inhibitory role of surface tension in elastic creasing \cite{amar2010swelling, haywood2010nucleation,mora2011surface} and cavitation \cite{gent1990cavitation}. Elastic analogues of other traditional fluid instabilities, including Saffman-Taylor fingering \cite{shull2000fingering, saintyves2013bulk,bigginsfingering,biggins2015fluid} and Rayleigh-Taylor fingering \cite{mora2014gravity,liang2015gravity}, have also recently been reported.

The small soft structures in which elastocapillary effects are important abound in biology.  For example, competition between elasticity and surface tension is well documented in pulmonary airways \cite{hazel2005surface}, which are only able to inflate when surface tension is reduced by pulmonary surfactant prior to birth \cite{avery1959surface}.  Furthermore, a growing number of biological organs, including  villi \cite{shyer2013villification, hannezo2011instabilities} and the mammalian brain \cite{toro2005morphogenetic, tallinen2014gyrification},  are now understood to be sculpted into their complex shapes by elastic instabilities \cite{BardRoss2,savin2011growth, li2011surface,tallinen2013surface, tallinen2015mechanics}. Since developing organs (and tumors) are small and soft, surface tension must also shape organs \cite{hannezo2012mechanical, dervaux2011shape}, making elastocapillary effects an essential component of a mechanical theory of morphogenesis. 

\begin{figure}[h]
\includegraphics[width=0.86\columnwidth]{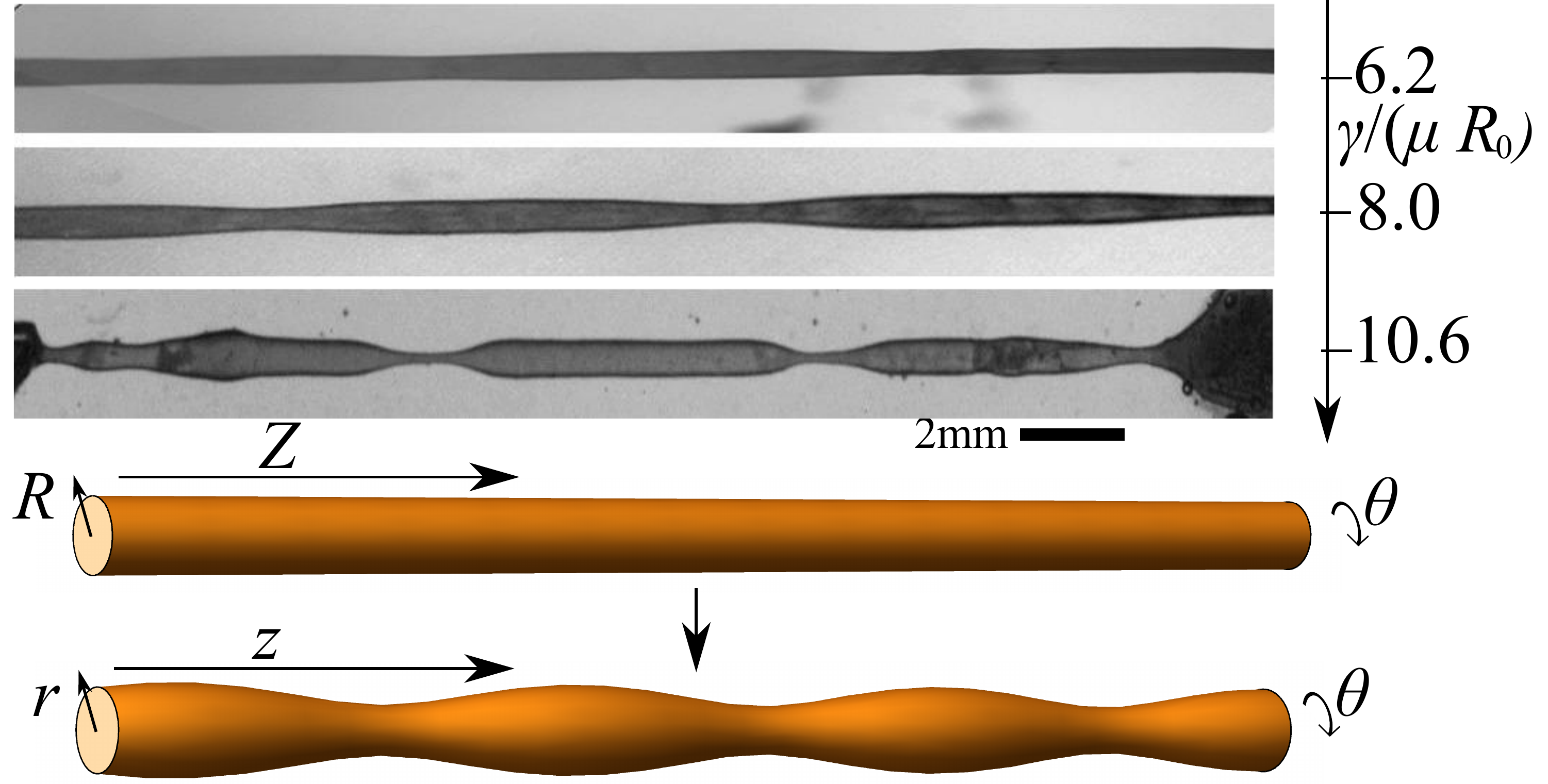}
\caption{Solid elastic cylinders undergo a peristaltic instability if surface tension, $\gamma$, is sufficient. Top: Experiment in gels from \cite{mora2010capillarity}. Bottom: Schematic of the instability.}\label{fig:mora_peri}
\end{figure}

The canonical example of surface tension driven instability in solids is the Rayleigh-Plateau instability in soft-solid cylinders \cite{barriere1996peristaltic, mora2010capillarity,ciarletta2012peristaltic,ciarlettanonlinearjmps, ciarlettanonlinearpre}, seen in fig.\ \ref{fig:mora_peri}. As in liquids, the instability originates in the geometric fact that a volume preserving peristaltic undulation with wavelength $\lambda$ along a cylinder of radius $R_0$ reduces its surface area provided  $\lambda>2 \pi R_0$. If $R_0 \lesssim l_{cap}$ the resultant saving in surface energy exceeds the  elastic energy, so the cylinder is unstable. Previous groups have observed this behavior \cite{mora2010capillarity}, predicted its onset \cite{barriere1996peristaltic,mora2010capillarity}, considered stretched cylinders \cite{ciarletta2012peristaltic} and calculated the high-amplitude behavior of the instability \cite{ciarlettanonlinearjmps, ciarlettanonlinearpre}. However, all these treatments contain a common surprise: the instability is governed by two length-scales, $R_0$ and $l_{cap}$, which are comparable around threshold, so one  expects $\lambda\sim R_0\sim l_{cap}$. Instead the first unstable mode has infinite wavelength.  We consider four geometries: a solid cylinder, an incompressible (e.g.\ fluid filled) cylindrical cavity in a bulk solid, a solid cylinder within a bulk solid, and a hollow cylindrical cavity in a bulk elastic solid. In the latter two cases, we discover peristaltic instabilities with $\lambda\sim l_{cap}$, demonstrating that the long-wavelength instability in cylinders is an anomaly rather than a signature of elastocapillary instabilities. Since all four systems are described by the same two length-scales, the finite $\lambda$ is not explicable as long-wavelength behavior trivially curtailed by a new length-scale (such as a finite length or finite outer radius), but rather is a reversion to the $\lambda \sim R_0 \sim l_{cap}$ behaviour one might have anticipated in all four cases. 

Our search for finite-wavelength solid Rayleigh-Plateau instabilities is further motivated by the existence of finite-wavelength instabilities in other areas of elasticity \cite{cerda2003geometry,mora2014gravity,bigginsfingering, dervaux2011buckling}. The full spectrum of wavelengths is seen in the compressive folding of a growing/swelling elastic layer adhered to an infinite elastic substrate \cite{cai1999imperfection, cao2012wrinkling, cao2012wrinkles}. If the layer is stiff, it folds with long but finite wavelength \cite{allen1969}, attaining the $\lambda\to \infty$ Euler Buckling limit when the substrate's modulus becomes negligible. In the opposite limit, if the layer has a modulus comparable to \cite{hutchinson_growth, tallinen2015mechanics} or much lower than \cite{amar2010swelling, haywood2010nucleation} the substrate's, the layer undergoes the zero-wavelength Biot surface instability \cite{biotbook}, leading to the formation of cusped folds known as creases or sulci \cite{hohlfeld2011unfolding, tallinen2013surface}.  These instabilities, like the solid Rayleigh-Plateau instabilities discussed here, are governed by a biharmonic (or similar) bulk equation augmented by boundary conditions at the interfaces. However, there is no obvious mapping between the sets of the problems, and the underlying physics is quite different --- compressive folding increases the boundary's area to release compression, whereas Rayleigh-Plateau instabilities are driven by area reduction --- so direct calculation in the Rayleigh-Plateau cases are necessary.

Beyond the issue of finite vs infinite wavelength, the final (compressible) cavity geometry explored here mimics channels through soft materials such as capillaries, airways, and those used in microfluidics. For this reason, we verify and extend this result by using finite elements to compute the high-amplitude behavior of this instability. We discover that it is strongly subcritical (like Rayleigh-Plateau in solid-cylinders \cite{ciarlettanonlinearpre}), with the channel jumping to an undulating state with very high strains and apparent channel closure. This suggests that experiencing the instability would be catastrophic for a fluid bearing channel's function, leading directly to blockage and/or material failure. Thus our result gives the lower limit for the radius of an un-reinforced channel through a solid before surface tension precipitates it's collapse.

%Some elastic-instabilities, such as wrinkling \cite{cerda2003geometry}, and the gravitational instability of an elastic slab \cite{mora2014gravity}, are intrinsically finite wavelength, meaning the first unstable mode has finite wavelength. Even in systems where the first instability has infinite wavelength, such as Euler-Bucking \cite{euler1759force} and the Rayleigh-Plateaux instability in solid cylinders, long but finite wavelength modes destabilize as loading is increased further \cite{barriere1996peristaltic}, allowing instability in finite length systems. 

\section{Instability in a solid cylinder}\label{sect:cylinder}
We first consider a long incompressible neo-Hookean cylinder with shear modulus $\mu$, surface tension $\gamma$ and length $L$, initially occupying $R<R_0$ in an $(R,\theta,Z)$ cylindrical coordinate system, as sketched in fig.\ \ref{fig:mora_peri}. For peristaltic deformations $(R,\theta,Z)$ is displaced to the final state coordinates $(r,\theta,z)$, so the deformation gradient is 
\begin{equation}
F=\left(\begin{matrix}
\frac{\partial r}{\partial R}\big|_{Z}& 0 &\frac{\partial r}{\partial Z}\big|_{R}\\
0 & r/R & 0\\
\frac{\partial z}{\partial R}\big|_{Z} & 0 & \frac{\partial z}{\partial Z}\big|_{R}
\end{matrix}\right).
\end{equation}
 The total elastic energy of the cylinder is given by
\begin{align}
E_{el}=&\int_{0}^{L}\int_0^{R_0} \frac{1}{2} \mu\left(\Tr{F F^T}-3\right)2\pi R\mathrm{d}R\mathrm{d}Z ,\label{eq:enel}\end{align}
 while the surface energy is given by
\begin{align}
E_{s}=\gamma \int_{0}^{L}2 \pi r(R_0,z) \sqrt{1+\left(\frac{\partial r(R_0,z)}{\partial z}\right)^2
}\mathrm{d}z.\label{eqn:surface_energy}
\end{align}
and  $\Det{F}=1$ encodes incompressibility. Following Ben-Amar \cite{ciarletta2012peristaltic} we represent the deformation in a mixed coordinates, specifying $r(R,z)$ and $Z(R,z)$. These mixed coordinates allow us to evaluate the final state surface energy $E_s$ directly without requiring $F^{-T}$ as one would working in normal fully-reference state coordinates. Furthermore, they also allow us to implement full rather than linearized incompressibility \cite{ciarletta2012peristaltic, carroll2007generating, ben2010swelling, ciarletta2012papillary, biggins2014exactly, biggins20142d}: in this representation the deformation gradient is
\be
F=\left(\begin{matrix}
\frac{\partial r}{\partial R}\big|_z-\frac{\partial r}{\partial z}\big|_R\frac{\partial Z}{\partial R}\big|_z\frac{\partial Z}{\partial z}\big|_R^{-1} & 0 & \frac{\partial r}{\partial z}\big|_R\frac{\partial Z}{\partial z}\big|_R^{-1}\\
0 & r/R & 0\\
-\frac{\partial Z}{\partial R}\big|_z\frac{\partial Z}{\partial z}\big|_R^{-1} & 0 & \frac{\partial Z}{\partial z}\big|_R^{-1}
\end{matrix}\right),
\ee
and, although this is rather more complicated than the standard representation, its determinant is simply $\det{F}=r \frac{\partial r}{\partial R}\big|_z/\left( R \frac{\partial Z}{\partial z}\big|_R\right)$, so we can enforce $\Det{F}=1$ by introducing the mixed-coordinate streamline function $\Phi(R,z)$ and setting
\be
Z=\frac{1}{R}\frac{\partial \Phi}{\partial R}\bigg|_z ,\ \ \ \  \ \ \ \ r^2=2\frac{\partial \Phi}{\partial z}\bigg|_R.\label{eq:streamline}
\ee
Setting $\Phi=\half R^2 z$ gives the undeformed state, so we examine the stability of this state to infinitesimal peristaltic distortions with wavenumber $k$ by considering
\be
\Phi=\half R^2z+\epsilon R f(R) \sin(k z)/k\label{eq:streamline linearized}
\ee
such that, to leading order, the cylinder adopts the profile $r(R_0,z)=R_0+\epsilon f(R_0) \cos(k z)$. Expanding $E_s$ to quadratic order in $\epsilon$ and conducting the $z$ integral yields the consequent  surface energy,
\begin{equation}
E_s=\gamma\left(2 \pi R_0 L+ \epsilon^2 \frac{\pi f(R_0)^2 L}  {2 R_0}\left(k^2 R_0^2-1\right)\right).\label{eqn:surface_peri}
\end{equation} 
We immediately see the geometric origin of the instability: the peristaltic shape change reduces the cylinder's area if $k<1/R_0$ and increases it if $k>1/R_0$, with longer wavelengths ($k\to0$) providing the biggest reduction.

Similarly, the linearized deformation gradient is
\begin{align}
%\left(
%\begin{array}{ccc}
% \alpha_{Rz}- \alpha_{z}/{R} & 0 & \alpha_{zz} \\
% 0 & \alpha_{z} & 0 \\
%\alpha_{R}/{R}-\alpha_{RR} & 0 & -\alpha_{Rz} \\
%\end{array}
%\right)
F\!=&I\!+\!\epsilon\!\left(\!\!\!
\begin{array}{ccc}
   f' \cos k z& 0 & -k f \sin k z \\
 0 & \frac{ f\! \cos k z }{R} & 0 \\
 \left(\!\!\frac{f}{R^2}\!-\!\frac{f'}{R}\!-\!f''\!\right)\! \frac{\sin k z}{k} & 0 & \!\!- \left(\!\!\frac{f}{R}\!+\!f'\!\right)\! \cos k z\!\! \\
\end{array}\!
\right)\!\!.\label{eqn:F_peri}
\end{align}
Substituting this into (\ref{eq:enel}), expanding to quadratic order and conducting the $z$ integral, (neglecting the higher-order difference between $\mathrm{d}Z$ and $\mathrm{d}z$) reveals the consequent elastic energy is:
\begin{align}
E_{el}\!&=\!\epsilon^2\!\!\int_{0}^{R_0}\!\!\! \frac{\pi L  \mu}{2 k^2 R^3}  \bigg[R^4 f''^2+2 R^3 f' f''+R^2 \left(4 k^2 R^2+1\right) f'^2 \notag \\ &\!2 R f \left[R(k^2R^2-1)f'\right]'+ \left(k^2 R^2+1\right)^2 f^2\bigg] \mathrm{d} R.
\end{align}

Minimizing $E_{el}+E_{s}$ with respect to variations of $f$ yields a biharmonic-like fourth-order elastic equation \cite{mora2010capillarity}
\be
\mathcal{L}^2[f]=0\ \textrm{ with }\ \mathcal{L}=\frac{\partial^2}{\partial R^2}+\frac{1}{R}\frac{\partial }{\partial R}-\frac{1}{R^2}-k^2,\label{eq:biharmonic}
\ee
with two natural boundary conditions, equivalent to imposing zero normal and shear stress on the boundary:
\begin{align}
&R_0^2  f'' +R_0 f'+  \left(k^2 R_0^2-1\right)f=0,\label{eq:BCcylinder1}\\
&R_0^3 f'''\!+3 R_0 ^2 f''\!\!-3k^2 R_0 ^3 f'\!=\!\frac{\gamma}{\mu R_0}k^2 R_0^2(k^2 R_0^2\!-\!1)f. \label{eq:BCcylinder2}
\end{align}
\begin{figure}[h]
\includegraphics[width=0.95\columnwidth]{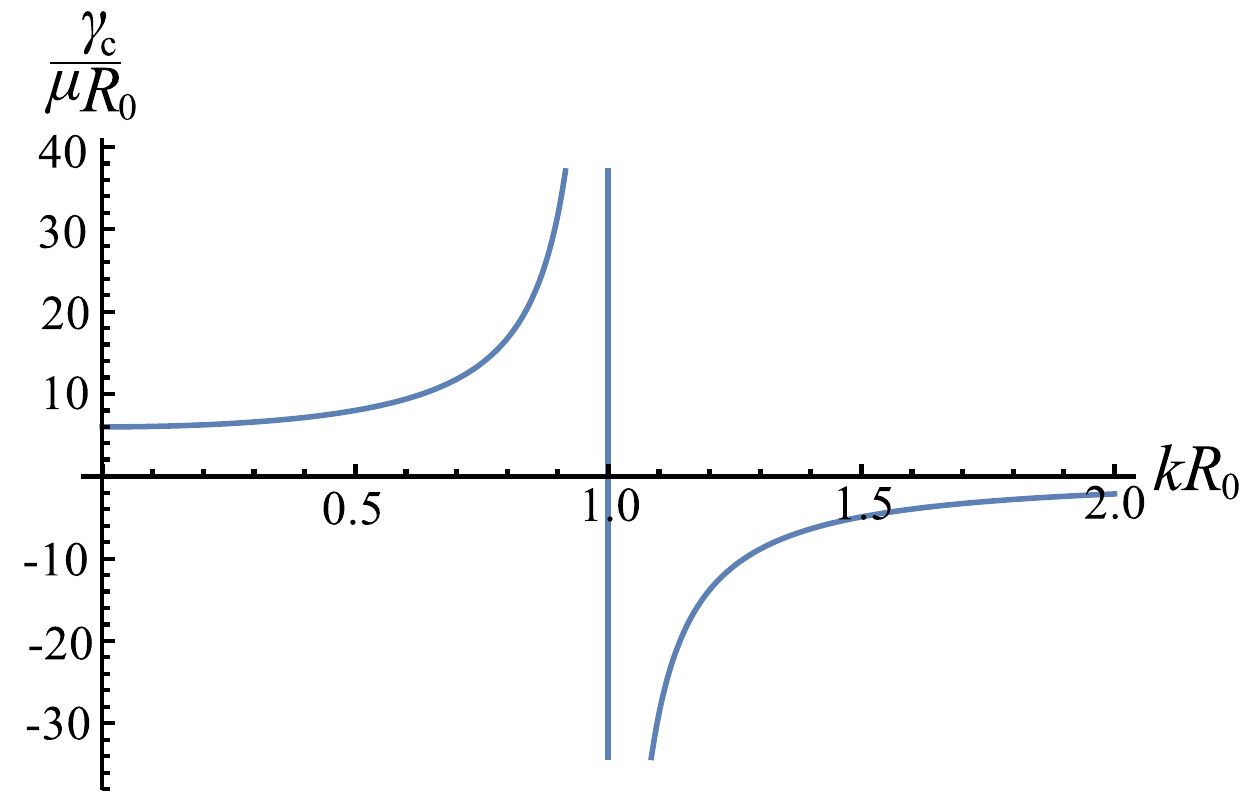}
\caption{Critical surface tension as a function of $k$ for a solid cylinder. The first unstable mode has infinite wavelength.}\label{fig:cylinder}
\end{figure}

The bulk equation, (\ref{eq:biharmonic}), is solved by a sum of four modified Bessel functions,
\be
f\!=c_1I_1(kR)+c_2kRI_2(kR)+c_3K_1(kR)+c_4kRK_2(kR),\label{eq:general solution}
\ee
but for a cylinder we require $c_3=c_4=0$, so that  $f(0)=f'(0)=0$ to ensure finite deformation ($F_{zR}$) at $R=0$. Imposing the first boundary condition requires
 \be
c_1=\left[1-\frac{kR_0I_0(kR_0)}{I_1(kR_0)}\right]c_2,
\ee
while the second b.c.\ yields the minimum $\gamma$ for instability:
\be
\gamma_c=\frac{2\mu R_0}{1-k^2R_0^2}\left[-(1+k^2R_0^2)+k^2R_0^2\frac{I_0(kR_0)^2}{I_1(kR_0)^2}\right].\label{eq:surface tension cylinder}
\ee
As always with linear-stability analysis, this solution marks the point where the $\epsilon^2$ term in the total energy is zero, segregating stable cases (positive coefficient) and unstable cases (negative coefficient). Fig.~\ref{fig:cylinder} shows this critical surface tension as a function of the wave number. As expected, the threshold surface tension diverges as $kR_0\to1$. In agreement with previous studies \cite{barriere1996peristaltic, mora2010capillarity, ciarletta2012peristaltic}, we find the long-wavelength ($k=0$) mode becomes unstable first at $\gamma_c=6\mu R_0$. The negative $\gamma_c$ for $kR_0>1$ indicates that instability only occurs with a sufficiently negative surface tension --- i.e. an energy reduction for creating surface --- impossible for a fluid, since such a fluid would simply vaporize,  but possible for an active solid such as a spontaneously folding epithelial sheet \cite{fleury2012coupling}. Far above threshold many modes are unstable, and previous authors\cite{barriere1996peristaltic} have argued that due to rate effects such as viscosity, the fastest growing mode may well have finite wavelength, much like in fluids. However, close to threshold such rate effects can only choose between the very-long wavelength modes that are unstable, so, at least in this region, the instability is unambiguously long wavelength.

\section{Incompressible cylindrical cavity}\label{sect:cavity}

\begin{figure}[h]
\includegraphics[width=0.9\columnwidth]{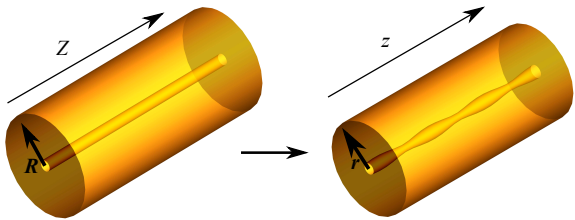}
\caption{Surface tension driven peristaltic instability in a cylindrical cavity, radius $R_0$, through an infinite elastic solid. }\label{fig:cylinder_cav_diagram}
\end{figure}

We next consider a cylindrical cavity, with radius $R_0$, running through an infinite incompressible elastic solid. The cavity's surface energy is still eqn.\ \ref{eqn:surface_energy} so, as in the solid-cylinder case, peristaltic undulations with $k<1/R_0$ will cause its area to decrease, implying sufficient surface tension will destabilize the cavity as sketched in fig.\ \ref{fig:cylinder_cav_diagram}. The elastic energy density is the same as before, but now integrated from $R_0$ to $\infty$. We first consider a cavity that preserves its volume, either by being filled with an incompressible fluid, or because the incompressible bulk solid is clamped at infinity. In this case, prior to instability there is no deformation so, at instability, the elastic fields must still satisfy eqn.\ \ref{eq:biharmonic}, and thus still be eqn.\ \ref{eq:general solution}. However the cavity geometry requires $c_1=c_2=0$, so that the displacement decays as $R\to \infty$, reducing the solution to 

\begin{figure}[h]
\includegraphics[width=0.95\columnwidth]{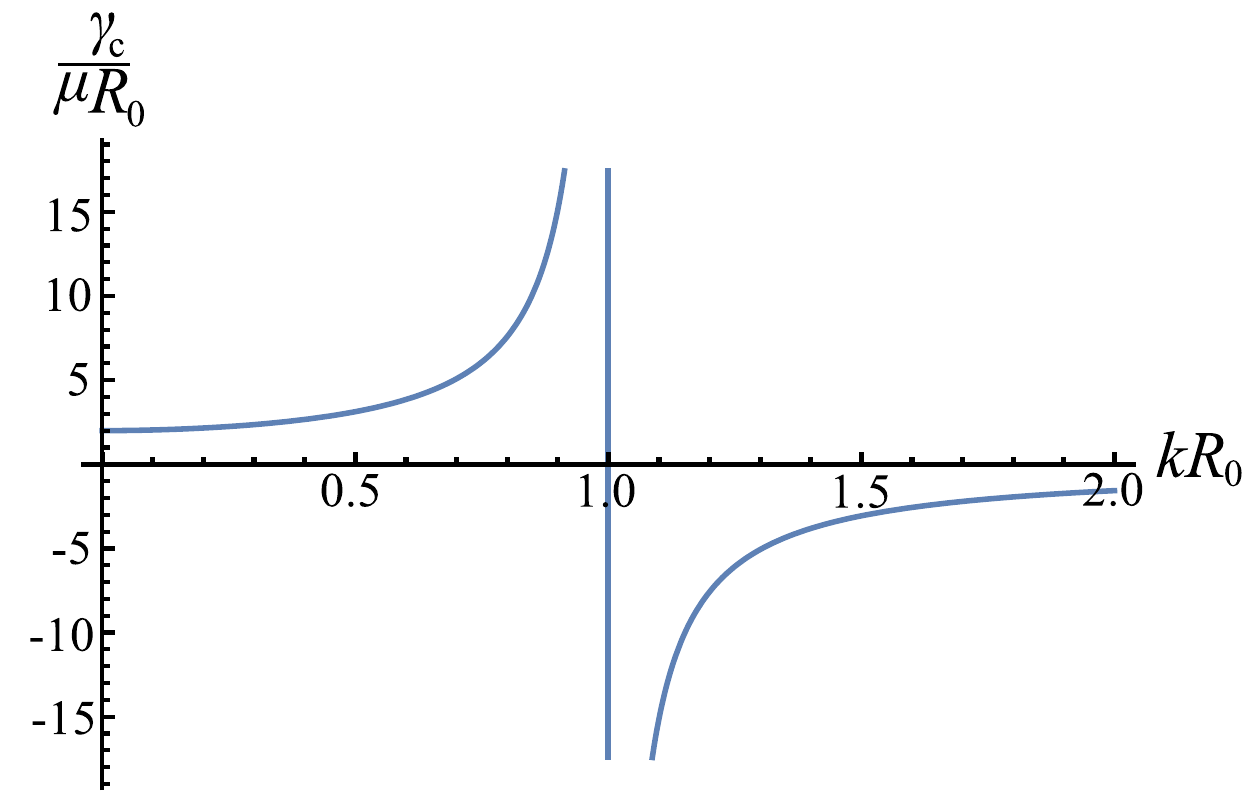}
\caption{Critical surface tension as a function of $k$ for an incompressible cylindrical cavity. The first unstable mode again has infinite wavelength.}\label{fig:cylinder_cav}
\label{fig:cavity}
\end{figure}

\be
f=c_3K_1(kR)+c_4kRK_2(kR).\label{eq:solution cavity}
\ee
The boundary conditions at $R=R_0$ are almost those in the cylindrical case (eqns.\ \ref{eq:BCcylinder1}-\ref{eq:BCcylinder2}), except that, since $R_0$ is  now the lower limit of the elastic energy integral the elastic contributions flip sign. Eqn.\ \ref{eq:BCcylinder1} simply flips sign entirely, and is thus unchanged, requiring:
\be
c_3=-\left[1+\frac{kR_0K_0(kR_0)}{K_1(kR_0)}\right]c_4.
\ee
However, the second boundary condition changes to
\be
-R_0^3 f'''\!-3 R_0 ^2 f''\!+3k^2 R_0 ^3 f'\!=\frac{\gamma}{\mu R_0}k^2 R_0^2(k^2 R_0^2-1)f, \label{eq:BC cav2}
\ee
from which we deduce the critical $\gamma$ for instability is
\be
\gamma_c=\frac{2\mu R_0}{1-k^2R_0^2}\left(1+k^2R_0^2-k^2R_0^2\frac{K_0(kR_0)^2}{K_1(kR_0)^2}\right).\label{eq:surface tension cavity}
\ee
We plot this critical surface tension as a function of wavelength in Fig.~\ref{fig:cavity}. As expected, sufficient surface tension will generate instability for any peristaltic undulation with $k<1/R_0$. As before the first unstable mode is the long wavelength $k\to0$ mode, but instability now occurs for  $\gamma_>2\mu R_0$, just one-third of the surface tension required to destabilize a solid cylinder.

\section{concentric cylinders}\label{sect:concentric cylinders}
We next consider a solid cylinder with modulus $\mu_i$ running through an infinite solid with modulus $\mu_o$. The two solids are materially connected, but the interface has a surface energy $\gamma$. Again the surface energy is eqn.\ \ref{eqn:surface_energy}, but the elastic energy now is the sum of contributions from  the cylinder and the bulk. Prior to instability incompressibility guarantees there is no displacement, so the elastic fields for the instability are once again described by eqn.\ \ref{eq:biharmonic}, and thus of the form
\begin{align}
f=\begin{cases} 
c_1I_1(kR)+c_2kRI_2(kR)& R< R_0 \\
c_3K_1(kR)+c_4kRK_2(kR)&  R>R_0. \\
   \end{cases}\end{align}
   \begin{figure*}
\begin{center}
\includegraphics[width=180mm]{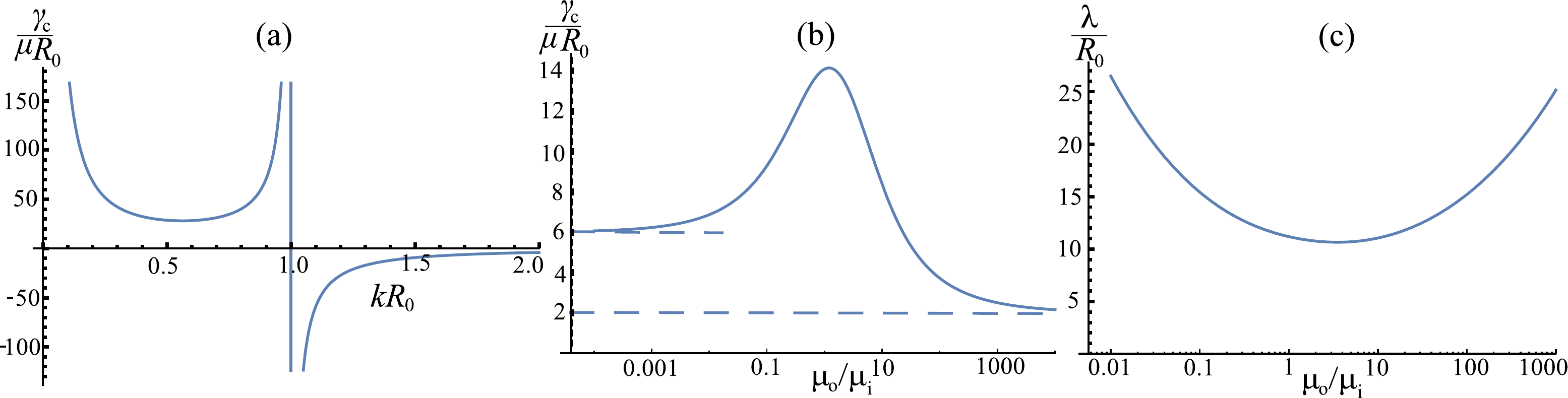}
\caption{Peristaltic instability in a solid cylinder of modulus $\mu_i$ through a bulk solid with modulus $\mu_o$. (a) Critical surface tension  for instability for concentric cylinders with $\mu_i=\mu_o$. The first unstable mode has finite wavelength with $k R_0=0.561992...$. (b) Surface tension for first instability as a function of modulus ratio. (c) First unstable wavelength as a function of modulus ratio. }
\label{fig:concentric-gamma}
\end{center}
\end{figure*}
Material continuity requires continuity of $Z(R,z)$ and $r(R,z)$, and hence  $\left[\!\left[f\right]\!\right]$=0 and $\left[\!\left[(Rf)'\right]\!\right]=0$, where we define $\left[\!\left[\Psi\right]\!\right]=\Psi(R_0^+)-\Psi(R_0^-)$ as the jump in $\Psi$ over the $R_0$ interface. In full these two conditions are
\begin{align}
c_1I_1(kR_0)\!+\!c_2kR_0I_2(kR_0)&\!=\!c_3K_1(kR_0)\!+\!c_4kR_0K_2(kR_0),\notag\\
c_1I_0(kR_0)\!+\!c_2kR_0I_1(kR_0)&\!=\!c_3K_0(kR_0)\!+\!c_4kR_0K_1(kR_0).\label{eq:compatibility}
\end{align}
Finally, the two natural boundary conditions, equivalent to requiring mechanical equilibrium in the radial and normal directions on the boundary, are
\begin{align}
&\left[\!\left[\mu\left( R^2  f'' +R f'+  \left(k^2 R^2-1\right)f\right)\right]\!\right]=0,\label{eq:BC_both}\\
& \left[\!\left[\mu\left(R^3 f'''\!+3 R ^2 f''\!\!-3k^2 R ^3 f'\right)\right]\!\right]=\!\frac{\gamma}{ R}k^2 R^2(k^2 R^2\!-\!1), \notag
\end{align}
where 
\begin{align}
\mu=\begin{cases} 
\mu_i& R< R_0 \\
\mu_o&  R>R_0. \\
   \end{cases}\end{align}
The four algebraic equations (\ref{eq:compatibility}-\ref{eq:BC_both}) are linear in the $c_i$ and $\gamma$, so it is easy to solve them for $c_1$, $c_2$, $c_3$ and the critical surface tension for instability, $\gamma_c$, as in the previous cases.  However, since the resulting expression is very long for arbitary $\mu_i$, $\mu_o$, we display here only the result for the special case, $\mu_i=\mu_o=\mu$, which gives:
\be
\gamma_c=\frac{2\mu}{k(k^2R_0^2-1)[I_0(kR_0)K_1(kR_0)-I_1(kR_0)K_2(kR_0)]}.
\ee

%\begin{figure}[h]
%\includegraphics[width=\columnwidth]{concentric-gamma.pdf}
%\caption{Critical surface tension  for instability for concentric cylinders with equal modulus. The first unstable mode has finite wavelength with $k R_0=0.561992...$ }\label{fig:concentric-gamma}
%\end{figure}
%
%\begin{figure}[h]
%\includegraphics[width=\columnwidth]{concentric-lambda-log.pdf}
%\caption{First unstable wavelength for concentric cylinders as a function of modulus ratio.}\label{fig:concentric-lambda}
%\end{figure}
%
%\begin{figure}[h]
%\includegraphics[width=\columnwidth]{concentric-gamma-log.pdf}
%\caption{Surface tension for first instability for concentric cylinders as a function of modulus ratio.}\label{fig:concentric-lambda-gamma}
%\end{figure}

As seen in fig.\ \ref{fig:concentric-gamma}a the surface tension required for instability diverges as  $kR_0\rightarrow1$, in accordance with the previous two cases. However, now the critical surface tension also diverges in the long-wavelength limit  $kR_0\rightarrow 0$, whereas in the previous two cases this limit is the first unstable mode. Rather, in the case of concentric cylinders with equal modulus, the first unstable mode has $k R_0=0.561992...$, and consequently a finite wavelength $\lambda\approx11 R_0$. The wavenumber and surface tension of the first unstable mode are shown as a function of modulus ratio in figs \ref{fig:concentric-gamma}b\&c. In general, the first unstable mode remains at finite wavelength provided both $\mu_i$ and $\mu_o$ are finite, and remains in the range $11 <\lambda/R_0<25 $ over  five magnitudes of modulus ratios.

\section{squeezable cavity}\label{sect:squeezable cavity}
Finally, we return to a long cylindrical cavity in a bulk elastic solid, as in section \ref{sect:cavity}, but no longer require that the cavity conserve its volume, instead imagining that it is completely empty, or filled with a fluid that can be squeezed out. In this case, even when surface tension is insufficient to cause a peristaltic instability, it will cause the radius of the cavity to decrease, reducing its surface area and its volume. There is only one deformation that is purely radial, independent of $z$ and volume conserving in the solid, $r=\sqrt{R^2-\alpha R_0^2}$ or, equivalently
\be
\Phi(R,z)=\half(R^2-\alpha R_0^2)z,
\ee
where the dimensionless parameter, $0<\alpha<1$, describes the degree of contraction, with $\alpha=1$ indicating the cavity has completely closed. This deformation leads to a surface energy
\be
E_{s}=\gamma L 2 \pi R_0  \sqrt{1-\alpha }
\ee
and an elastic energy
\be
E_{el}=\mu  \pi  R_0^2 L \alpha   \log \left(1-\alpha\right)^{-1/2}.
\ee
Minimizing $E_{el}+E_{s}$ with respect to $\alpha$, we find that the surface tension required to produce a given $\alpha$ is
\be
\frac{\gamma}{\mu R_0}=\frac{\alpha -(1-\alpha ) \log (1-\alpha )}{2 \sqrt{1-\alpha }},\label{eq:C-gamma}
\ee
a continuous and monotonic relationship, with $\alpha=0$ when there is no surface tension, and $\alpha\to1$ (i.e. full closure of the cavity) as $\gamma \to \infty$.

\begin{figure*}
\begin{center}
\includegraphics[width=180mm]{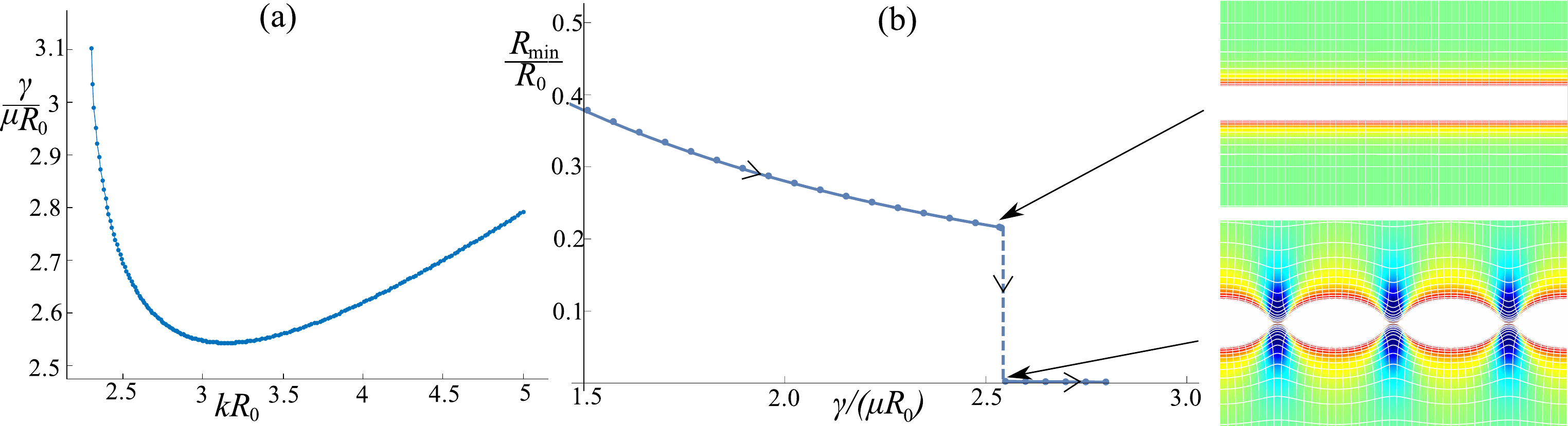}
\caption{(a) Critical surface tension as a function of the axial wave number for a squeezed cavity: the threshold surface tension is minimized at  $\gamma_c\approx 2.543... \mu R_0$ by  $kR_0=3.145...$. (b) Finite element analysis of the squeezed cavity instability. The minimum (final state) radius of the cavity, $R_{min}$, falls as $\gamma$ is increased. As shown in the insets on the right, for $\gamma$ below threshold, the radius contracts homogeneously along the whole channel as described by eqn.\ \ref{eq:C-gamma}. At threshold, the instability proceeds sub-critically and the system jumps to a highly peristaltic state with a zero (or almost zero) minimum radius. Inset color indicates isotropic stress (dark red $<-1.25\mu$, dark blue $>1.25\mu$) and the lines indicate every 8th mesh line.
 }
\label{fig:squeeze axial}
\end{center}
\end{figure*}

We next add an infinitesimal peristaltic perturbation to this base state:
\be
\Phi(R,z)=\half(R^2-\alpha R_0^2)z+\epsilon g(R) \sin(k z),
\ee
which generates the following  leading order corrections to the surface energy:
\begin{equation}
\delta E_s=\epsilon ^2\frac{ \gamma L  \pi    k^2 g(R_0)^2 }{2 \left((1-\alpha )^{3/2} R_0^3\right)}\left((1-\alpha ) k^2 R_0^2-1\right).
\end{equation}
This correction is negative (i.e.\ the perturbation reduces area) if  $k<1/(R_0\sqrt{1-\alpha})$ since the base deformation leads to a reduction of the cavity's radius to $R_0\sqrt{1-\alpha}$. With the aid of a computer algebra package, it is also possible to evaluate the correction to the elastic energy:
\begin{align}
\delta E_{el}\!&=\!\int_{R_0}^{\infty}\!\!\!\epsilon^2\frac{\pi L  \mu}{2 R^3 r^6}  \left(R^2 r^6 g''^2+r^4 g'^2 \left(k^2 R^2 \left(3 R^2+r^2\right)+r^2\right)\right.\notag\\&\left.+2 k^2 r^2 g \left(R^4 r^2 g''+R \left(-3 R^4-R^2 r^2+r^4\right) g'\right)\right.\notag\\&\left.-2 R r^6 g' g''+g^2 \left(k^4 R^4 r^4+4 k^2 R^6\right)\right),
\end{align}
where we use $r=\sqrt{R^2-R_0^2 \alpha}$ to denote the base-state radius of material originally at $R$. Minimizing the total energy variationally with respect to $g(R)$ requires
\begin{align}
&R^3 r^4 g^{(4)}-2 R^2 r^4 g^{(3)}-R r^2 g'' \left(k^2 R^2 \left(R^2+r^2\right)-3 r^2\right)\notag \\&+g' \left(k^2 R^2 \left(2 R^4-R^2 r^2+r^4\right)-3 r^4\right)\notag \\&+k^2 R g \left(R^4 \left(k^2 r^2-2\right)+2 r^4\right)=0
\end{align}
in the bulk and the natural boundary conditions
\begin{align}
(\alpha -1) \left(g'-R_0 g''\right)+k^2 R_0 g&=0
\end{align}
\begin{align}
&\left(\alpha ^2-2\right) k^2 R_0^2 g+(\alpha -1)  \left((\alpha -3) k^2 R_0^2+1-\alpha\right)g'
\notag\\ &+(\alpha -1)^2 R_0 \left(g''-R_0 g^{(3)}\right)\notag\\&=\frac{\gamma }{\mu R_0}\sqrt{1-\alpha } k^2 R_0^2 \left((\alpha -1) k^2 R_0^2+1\right) g\end{align}
on the inner $R_0$ surface. The final two boundary conditions are simply that $g(R)$ and $g'(R)$ tend to zero as $R\to\infty$, so the displacements vanish at infinity. The above system of equations do not admit an analytic solution, but we are able to solve them numerically using the MATLAB routine bvp4c, to find the threshold surface tension required to trigger instability at each wavelength, which we plot in fig.\ \ref{fig:squeeze axial}a. Minimizing over wavelengths, we see that the first unstable mode has $kR_0=3.145...$ and is triggered when $\gamma\approx 2.543... \mu R_0$. 

Since this geometry directly mimics a micro-fluidic channel through a soft solid, we verify and extend our threshold result via cylindrically-symmetric large-strain finite-element simulation (details in appendix \ref{numerical_SI}) Figure \ref{fig:squeeze axial}b shows the minimum radius of the channel, $R_{min}$,  as surface tension is increased. Before threshold the channel contracts homogeneously so it's radius ($R_{min}$) falls but the channel remains cylindrical. The simulation points are in excellent agreement with the solid theoretical curve derived from eqn.\ \ref{eq:C-gamma}.  At threshold, the instability occurs with the expected wavelength and $\gamma$, and the channel jumps sub-critically to a highly deformed peristaltic state with a negligible minimum radius, effectively closing the channel. In our simulations the peristaltic deformation closed the channel to within $ 0.01 R_0$, a length-scale commensurate with  our mesh, and generated compressive stresses  $>100\mu$ at the inner radius. This suggests that, within the neo-Hookean cylindrically symmetric framework, the channel probably closes completely; this is plausible since the inward capillary stress, $\gamma/R$, grows as material moves inward, whilst the neo-Hookean constitutive law has no strain-stiffening behavior. Of course, in real materials divergent elastic stresses are impossible, so real channels will avoid them by plastic deformation, fracture, strain-stiffening or breaking cylindrical symmetry.  However, suffering the instability is assured to be a dramatic sub-critical event leading to almost or complete channel closure, so we propose that $R_0=0.393...\gamma /\mu$ is the smallest possible radius for an unreinforced open channel through an elastic solid.

\section{Conclusions}\label{sect:conclusions}
The main result of this paper is that the infinite wavelength surface tension driven instability of a soft incompressible cylinder is an anomaly; we should normally expect such instabilities in soft solids to have, at threshold, wavelengths $\lambda\sim l_{cap}$. To demonstrate this, we have calculated the threshold wavelength for two new geometries with cylindrical symmetry, a solid cylinder embedded in a bulk solid, and a cylindrical cavity through a bulk neo-Hookean solid, and shown that, in both cases, the instability
does indeed proceed with $\lambda\sim l_{cap}$. We anticipate these geometries will find direct relevance in biology since filaments and pipes through soft-solid tissues abound in physiology, are sculpted during development and fail during disease. 

There remains the questions of why the simple cylinder and the incompressible cylindrical cavity produce infinite wavelength instabilities. Looking at the surface energy for a peristaltic distortion, eqn.\ \ref{eqn:surface_peri}, we see that surface tension certainly favors long wavelength instabilities: the smaller $k$ is, the more surface energy is reduced for a given amplitude. We would normally expect distortions with infinite or vanishing wavelength to generate infinite shears, and hence to be suppressed elastically, and indeed if we look at the form of the peristaltic deformation gradient, eqn.\ \ref{eqn:F_peri}, we see that the  $r-z$ component diverges at short wavelengths, and the $z-r$ component diverges at long wavelengths, so we generically expect an intermediate wavelength to be preferred. However, the $z-r$ shears can also be suppressed, even at long wavelength and finite amplitude, if  $\frac{f}{R^2}\!-\!\frac{f'}{R}\!-\!f''=0$, requiring
\begin{equation}
f(R)=c_1 \frac{R}{R_0}+c_2 \frac{R_0}{R},
\end{equation}
and hence $\mathbf{\hat{z}}$ displacements 
\begin{equation}
u_z\equiv z-Z=-\epsilon\frac{2 c_1 }{k R_0} \sin (k z).
\end{equation}
The first of these solutions ($c_1$) generates the long wavelength solution for the simple cylinder, and the second ($c_2$) generates the analogous solution for the incompressible cavity. The solutions have markedly different character. The $c_2$ cavity solution has $Z=z$ and is thus a plane-strain solution associated with a radial displacements of the form $u_R\sim1/R$, which do decay as $R\to \infty$ but are nevertheless able to bring volume in from infinity since $R u_R$ remains finite. The $c_1$ solid cylinder solution is associated with diverging $z$ displacements (though not shears) as $k\to0$, which are generated because the cylinder breaks into  long regions which alternate between getting thinner and longer and getting fatter and shorter, generating big longitudinal displacements. Clearly these two solutions, with such different $u_z$, are not kinematically compatible with each other, so it is no surprise that when we consider a solid cylinder through a bulk solid these solutions are not accessed and the instability has finite wavelength. 

More generally, a full solution to a fourth-order elasticity problem requires four boundary conditions, and thus a solution with four constants of integration. Here we only have two compatible with long wavelength instabilities, so we generically expect surface tension instabilities, even in cylindrical geometries, to not be described by this simple form, and thus have finite wavelength.

{\it Acknowledgements}. C.X. thanks the China Scholarship Council for funding and  J.B thanks Trinity Hall and the 1851 Royal Commission for funding. Both thank T. Tallinen, whose code the finite element simulations in this paper are based on.

\appendix

\section{Details of numerical simulations}\label{numerical_SI}
Our simulations use an explicit finite-element method, based on the same code used in \cite{tallinen2014gyrification} and \cite{tallinen2015mechanics}. Since the problem has cylindrical symmetry, we construct the body from constant-strain triangular elements in the $r-z$ plane, each of which represents a triangularly cross-sectioned torus of the elastic body. The triangles form a rectangular mesh spanning from the inner radius, $R_0$, to an outer radius of $30 R_0$ and from $z=0$ to $z=\lambda=1.99... R_0$, with periodicity enforced in the $z$ direction. The mesh contains around 21000 nodes, and spans $z$ in 121 equally spaced segments, while in the radial direction it coarsens from a spacing of $R_0/6400$ at the inner boundary to a spacing of $1.5R_0$ at the outer boundary. Each triangle is assigned a compressible neo-Hookean elastic energy (within a quasi-incompressible nodal pressure formulation) with a bulk modulus $K = 2\times10^3\mu$. The force on each node is calculated as the gradient of the total energy with respect to nodal position, including surface energy at the inner surface, and the nodes are moved according to damped Newtonian dynamics. Surface tension was  increased sufficiently slowly as to be quasi-static, so although the simulation uses Newtonian dynamics, the states reported are converged energy minima. The good agreement between the predicted and observed threshold $\gamma$ (error in threshold $<0.2\%$) for instability suggests the outer radius is large enough, the bulk modulus high enough, and the mesh fine enough to mimic an incompressible infinite continuum material. The extremely fine mesh at the inner boundary is required to capture the extreme strain gradients at the inner points of the peristaltic state.



% Produces the bibliography via BibTeX.
%paste in by hand (from photo_eclipsing.bbl) the prepared bibliography when ready and comment out the line

\begin{thebibliography}{52}%
\makeatletter
\providecommand \@ifxundefined [1]{%
 \@ifx{#1\undefined}
}%
\providecommand \@ifnum [1]{%
 \ifnum #1\expandafter \@firstoftwo
 \else \expandafter \@secondoftwo
 \fi
}%
\providecommand \@ifx [1]{%
 \ifx #1\expandafter \@firstoftwo
 \else \expandafter \@secondoftwo
 \fi
}%
\providecommand \natexlab [1]{#1}%
\providecommand \enquote  [1]{``#1''}%
\providecommand \bibnamefont  [1]{#1}%
\providecommand \bibfnamefont [1]{#1}%
\providecommand \citenamefont [1]{#1}%
\providecommand \href@noop [0]{\@secondoftwo}%
\providecommand \href [0]{\begingroup \@sanitize@url \@href}%
\providecommand \@href[1]{\@@startlink{#1}\@@href}%
\providecommand \@@href[1]{\endgroup#1\@@endlink}%
\providecommand \@sanitize@url [0]{\catcode `\\12\catcode `\$12\catcode
  `\&12\catcode `\#12\catcode `\^12\catcode `\_12\catcode `\%12\relax}%
\providecommand \@@startlink[1]{}%
\providecommand \@@endlink[0]{}%
\providecommand \url  [0]{\begingroup\@sanitize@url \@url }%
\providecommand \@url [1]{\endgroup\@href {#1}{\urlprefix }}%
\providecommand \urlprefix  [0]{URL }%
\providecommand \Eprint [0]{\href }%
\providecommand \doibase [0]{http://dx.doi.org/}%
\providecommand \selectlanguage [0]{\@gobble}%
\providecommand \bibinfo  [0]{\@secondoftwo}%
\providecommand \bibfield  [0]{\@secondoftwo}%
\providecommand \translation [1]{[#1]}%
\providecommand \BibitemOpen [0]{}%
\providecommand \bibitemStop [0]{}%
\providecommand \bibitemNoStop [0]{.\EOS\space}%
\providecommand \EOS [0]{\spacefactor3000\relax}%
\providecommand \BibitemShut  [1]{\csname bibitem#1\endcsname}%
\let\auto@bib@innerbib\@empty
%</preamble>
\bibitem [{\citenamefont {Strutt}\ and\ \citenamefont
  {Rayleigh}(1879)}]{strutt1879instability}%
  \BibitemOpen
  \bibfield  {author} {\bibinfo {author} {\bibfnamefont {J.~W.}\ \bibnamefont
  {Strutt}}\ and\ \bibinfo {author} {\bibfnamefont {L.}~\bibnamefont
  {Rayleigh}},\ }\href@noop {} {\enquote {\bibinfo {title} {On the instability
  of jets},}\ } (\bibinfo {year} {1879})\BibitemShut {NoStop}%
\bibitem [{\citenamefont {Roman}\ and\ \citenamefont
  {Bico}(2010)}]{roman2010elasto}%
  \BibitemOpen
  \bibfield  {author} {\bibinfo {author} {\bibfnamefont {B.}~\bibnamefont
  {Roman}}\ and\ \bibinfo {author} {\bibfnamefont {J.}~\bibnamefont {Bico}},\
  }\href@noop {} {\bibfield  {journal} {\bibinfo  {journal} {Journal of
  Physics: Condensed Matter}\ }\textbf {\bibinfo {volume} {22}},\ \bibinfo
  {pages} {493101} (\bibinfo {year} {2010})}\BibitemShut {NoStop}%
\bibitem [{\citenamefont {Matsuo}\ and\ \citenamefont
  {Tanaka}(1992)}]{matsuo1992patterns}%
  \BibitemOpen
  \bibfield  {author} {\bibinfo {author} {\bibfnamefont {E.~S.}\ \bibnamefont
  {Matsuo}}\ and\ \bibinfo {author} {\bibfnamefont {T.}~\bibnamefont
  {Tanaka}},\ }\href {\doibase 10.1038/358482a0} {\bibfield  {journal}
  {\bibinfo  {journal} {Nature}\ }\textbf {\bibinfo {volume} {358}},\ \bibinfo
  {pages} {482} (\bibinfo {year} {1992})}\BibitemShut {NoStop}%
\bibitem [{\citenamefont {Barriere}\ \emph {et~al.}(1996)\citenamefont
  {Barriere}, \citenamefont {Sekimoto},\ and\ \citenamefont
  {Leibler}}]{barriere1996peristaltic}%
  \BibitemOpen
  \bibfield  {author} {\bibinfo {author} {\bibfnamefont {B.}~\bibnamefont
  {Barriere}}, \bibinfo {author} {\bibfnamefont {K.}~\bibnamefont {Sekimoto}},
  \ and\ \bibinfo {author} {\bibfnamefont {L.}~\bibnamefont {Leibler}},\
  }\href@noop {} {\bibfield  {journal} {\bibinfo  {journal} {The Journal of
  chemical physics}\ }\textbf {\bibinfo {volume} {105}},\ \bibinfo {pages}
  {1735} (\bibinfo {year} {1996})}\BibitemShut {NoStop}%
\bibitem [{\citenamefont {Mora}\ \emph {et~al.}(2010)\citenamefont {Mora},
  \citenamefont {Phou}, \citenamefont {Fromental}, \citenamefont {Pismen},\
  and\ \citenamefont {Pomeau}}]{mora2010capillarity}%
  \BibitemOpen
  \bibfield  {author} {\bibinfo {author} {\bibfnamefont {S.}~\bibnamefont
  {Mora}}, \bibinfo {author} {\bibfnamefont {T.}~\bibnamefont {Phou}}, \bibinfo
  {author} {\bibfnamefont {J.-M.}\ \bibnamefont {Fromental}}, \bibinfo {author}
  {\bibfnamefont {L.~M.}\ \bibnamefont {Pismen}}, \ and\ \bibinfo {author}
  {\bibfnamefont {Y.}~\bibnamefont {Pomeau}},\ }\href@noop {} {\bibfield
  {journal} {\bibinfo  {journal} {Physical review letters}\ }\textbf {\bibinfo
  {volume} {105}},\ \bibinfo {pages} {214301} (\bibinfo {year}
  {2010})}\BibitemShut {NoStop}%
\bibitem [{\citenamefont {Ciarletta}\ and\ \citenamefont
  {Ben~Amar}(2012{\natexlab{a}})}]{ciarletta2012peristaltic}%
  \BibitemOpen
  \bibfield  {author} {\bibinfo {author} {\bibfnamefont {P.}~\bibnamefont
  {Ciarletta}}\ and\ \bibinfo {author} {\bibfnamefont {M.}~\bibnamefont
  {Ben~Amar}},\ }\href@noop {} {\bibfield  {journal} {\bibinfo  {journal} {Soft
  Matter}\ }\textbf {\bibinfo {volume} {8}},\ \bibinfo {pages} {1760} (\bibinfo
  {year} {2012}{\natexlab{a}})}\BibitemShut {NoStop}%
\bibitem [{\citenamefont {Taffetani}\ and\ \citenamefont
  {Ciarletta}(2015{\natexlab{a}})}]{ciarlettanonlinearjmps}%
  \BibitemOpen
  \bibfield  {author} {\bibinfo {author} {\bibfnamefont {M.}~\bibnamefont
  {Taffetani}}\ and\ \bibinfo {author} {\bibfnamefont {P.}~\bibnamefont
  {Ciarletta}},\ }\href@noop {} {\bibfield  {journal} {\bibinfo  {journal}
  {Journal of the Mechanics and Physics of Solids}\ } (\bibinfo {year}
  {2015}{\natexlab{a}})}\BibitemShut {NoStop}%
\bibitem [{\citenamefont {Taffetani}\ and\ \citenamefont
  {Ciarletta}(2015{\natexlab{b}})}]{ciarlettanonlinearpre}%
  \BibitemOpen
  \bibfield  {author} {\bibinfo {author} {\bibfnamefont {M.}~\bibnamefont
  {Taffetani}}\ and\ \bibinfo {author} {\bibfnamefont {P.}~\bibnamefont
  {Ciarletta}},\ }\href {\doibase 10.1103/PhysRevE.91.032413} {\bibfield
  {journal} {\bibinfo  {journal} {Phys. Rev. E}\ }\textbf {\bibinfo {volume}
  {91}},\ \bibinfo {pages} {032413} (\bibinfo {year}
  {2015}{\natexlab{b}})}\BibitemShut {NoStop}%
\bibitem [{\citenamefont {Style}\ \emph
  {et~al.}(2013{\natexlab{a}})\citenamefont {Style}, \citenamefont
  {Boltyanskiy}, \citenamefont {Che}, \citenamefont {Wettlaufer}, \citenamefont
  {Wilen},\ and\ \citenamefont {Dufresne}}]{style2013universal}%
  \BibitemOpen
  \bibfield  {author} {\bibinfo {author} {\bibfnamefont {R.~W.}\ \bibnamefont
  {Style}}, \bibinfo {author} {\bibfnamefont {R.}~\bibnamefont {Boltyanskiy}},
  \bibinfo {author} {\bibfnamefont {Y.}~\bibnamefont {Che}}, \bibinfo {author}
  {\bibfnamefont {J.}~\bibnamefont {Wettlaufer}}, \bibinfo {author}
  {\bibfnamefont {L.~A.}\ \bibnamefont {Wilen}}, \ and\ \bibinfo {author}
  {\bibfnamefont {E.~R.}\ \bibnamefont {Dufresne}},\ }\href@noop {} {\bibfield
  {journal} {\bibinfo  {journal} {Phys. Rev. Lett.}\ }\textbf {\bibinfo
  {volume} {110}},\ \bibinfo {pages} {066103} (\bibinfo {year}
  {2013}{\natexlab{a}})}\BibitemShut {NoStop}%
\bibitem [{\citenamefont {Style}\ \emph
  {et~al.}(2013{\natexlab{b}})\citenamefont {Style}, \citenamefont {Hyland},
  \citenamefont {Boltyanskiy}, \citenamefont {Wettlaufer},\ and\ \citenamefont
  {Dufresne}}]{style2013surface}%
  \BibitemOpen
  \bibfield  {author} {\bibinfo {author} {\bibfnamefont {R.~W.}\ \bibnamefont
  {Style}}, \bibinfo {author} {\bibfnamefont {C.}~\bibnamefont {Hyland}},
  \bibinfo {author} {\bibfnamefont {R.}~\bibnamefont {Boltyanskiy}}, \bibinfo
  {author} {\bibfnamefont {J.~S.}\ \bibnamefont {Wettlaufer}}, \ and\ \bibinfo
  {author} {\bibfnamefont {E.~R.}\ \bibnamefont {Dufresne}},\ }\href@noop {}
  {\bibfield  {journal} {\bibinfo  {journal} {Nature Com.}\ }\textbf {\bibinfo
  {volume} {4}} (\bibinfo {year} {2013}{\natexlab{b}})}\BibitemShut {NoStop}%
\bibitem [{\citenamefont {Style}\ \emph
  {et~al.}(2013{\natexlab{c}})\citenamefont {Style}, \citenamefont {Che},
  \citenamefont {Park}, \citenamefont {Weon}, \citenamefont {Je}, \citenamefont
  {Hyland}, \citenamefont {German}, \citenamefont {Power}, \citenamefont
  {Wilen}, \citenamefont {Wettlaufer} \emph {et~al.}}]{style2013patterning}%
  \BibitemOpen
  \bibfield  {author} {\bibinfo {author} {\bibfnamefont {R.~W.}\ \bibnamefont
  {Style}}, \bibinfo {author} {\bibfnamefont {Y.}~\bibnamefont {Che}}, \bibinfo
  {author} {\bibfnamefont {S.~J.}\ \bibnamefont {Park}}, \bibinfo {author}
  {\bibfnamefont {B.~M.}\ \bibnamefont {Weon}}, \bibinfo {author}
  {\bibfnamefont {J.~H.}\ \bibnamefont {Je}}, \bibinfo {author} {\bibfnamefont
  {C.}~\bibnamefont {Hyland}}, \bibinfo {author} {\bibfnamefont {G.~K.}\
  \bibnamefont {German}}, \bibinfo {author} {\bibfnamefont {M.~P.}\
  \bibnamefont {Power}}, \bibinfo {author} {\bibfnamefont {L.~A.}\ \bibnamefont
  {Wilen}}, \bibinfo {author} {\bibfnamefont {J.~S.}\ \bibnamefont
  {Wettlaufer}},  \emph {et~al.},\ }\href@noop {} {\bibfield  {journal}
  {\bibinfo  {journal} {Proc. of the Nat. Acad of Sci.}\ }\textbf {\bibinfo
  {volume} {110}},\ \bibinfo {pages} {12541} (\bibinfo {year}
  {2013}{\natexlab{c}})}\BibitemShut {NoStop}%
\bibitem [{\citenamefont {Bico}\ \emph {et~al.}(2004)\citenamefont {Bico},
  \citenamefont {Roman}, \citenamefont {Moulin},\ and\ \citenamefont
  {Boudaoud}}]{bico2004adhesion}%
  \BibitemOpen
  \bibfield  {author} {\bibinfo {author} {\bibfnamefont {J.}~\bibnamefont
  {Bico}}, \bibinfo {author} {\bibfnamefont {B.}~\bibnamefont {Roman}},
  \bibinfo {author} {\bibfnamefont {L.}~\bibnamefont {Moulin}}, \ and\ \bibinfo
  {author} {\bibfnamefont {A.}~\bibnamefont {Boudaoud}},\ }\href@noop {}
  {\bibfield  {journal} {\bibinfo  {journal} {Nature}\ }\textbf {\bibinfo
  {volume} {432}},\ \bibinfo {pages} {690} (\bibinfo {year}
  {2004})}\BibitemShut {NoStop}%
\bibitem [{\citenamefont {Kim}\ and\ \citenamefont
  {Mahadevan}(2006)}]{kim2006capillary}%
  \BibitemOpen
  \bibfield  {author} {\bibinfo {author} {\bibfnamefont {H.-Y.}\ \bibnamefont
  {Kim}}\ and\ \bibinfo {author} {\bibfnamefont {L.}~\bibnamefont
  {Mahadevan}},\ }\href@noop {} {\bibfield  {journal} {\bibinfo  {journal}
  {Journal of Fluid mechanics}\ }\textbf {\bibinfo {volume} {548}},\ \bibinfo
  {pages} {141} (\bibinfo {year} {2006})}\BibitemShut {NoStop}%
\bibitem [{\citenamefont {Mora}\ \emph {et~al.}(2013)\citenamefont {Mora},
  \citenamefont {Maurini}, \citenamefont {Phou}, \citenamefont {Fromental},
  \citenamefont {Audoly},\ and\ \citenamefont {Pomeau}}]{mora2013solid}%
  \BibitemOpen
  \bibfield  {author} {\bibinfo {author} {\bibfnamefont {S.}~\bibnamefont
  {Mora}}, \bibinfo {author} {\bibfnamefont {C.}~\bibnamefont {Maurini}},
  \bibinfo {author} {\bibfnamefont {T.}~\bibnamefont {Phou}}, \bibinfo {author}
  {\bibfnamefont {J.-M.}\ \bibnamefont {Fromental}}, \bibinfo {author}
  {\bibfnamefont {B.}~\bibnamefont {Audoly}}, \ and\ \bibinfo {author}
  {\bibfnamefont {Y.}~\bibnamefont {Pomeau}},\ }\href@noop {} {\bibfield
  {journal} {\bibinfo  {journal} {Phys. Rev. Lett.}\ }\textbf {\bibinfo
  {volume} {111}},\ \bibinfo {pages} {114301} (\bibinfo {year}
  {2013})}\BibitemShut {NoStop}%
\bibitem [{\citenamefont {Amar}\ and\ \citenamefont
  {Ciarletta}(2010)}]{amar2010swelling}%
  \BibitemOpen
  \bibfield  {author} {\bibinfo {author} {\bibfnamefont {M.~B.}\ \bibnamefont
  {Amar}}\ and\ \bibinfo {author} {\bibfnamefont {P.}~\bibnamefont
  {Ciarletta}},\ }\href@noop {} {\bibfield  {journal} {\bibinfo  {journal}
  {Journal of the Mechanics and Physics of Solids}\ }\textbf {\bibinfo {volume}
  {58}},\ \bibinfo {pages} {935} (\bibinfo {year} {2010})}\BibitemShut
  {NoStop}%
\bibitem [{\citenamefont {Yoon}\ \emph {et~al.}(2010)\citenamefont {Yoon},
  \citenamefont {Kim},\ and\ \citenamefont {Hayward}}]{haywood2010nucleation}%
  \BibitemOpen
  \bibfield  {author} {\bibinfo {author} {\bibfnamefont {J.}~\bibnamefont
  {Yoon}}, \bibinfo {author} {\bibfnamefont {J.}~\bibnamefont {Kim}}, \ and\
  \bibinfo {author} {\bibfnamefont {R.~C.}\ \bibnamefont {Hayward}},\
  }\href@noop {} {\bibfield  {journal} {\bibinfo  {journal} {Soft Matter}\
  }\textbf {\bibinfo {volume} {6}},\ \bibinfo {pages} {5807} (\bibinfo {year}
  {2010})}\BibitemShut {NoStop}%
\bibitem [{\citenamefont {Mora}\ \emph {et~al.}(2011)\citenamefont {Mora},
  \citenamefont {Abkarian}, \citenamefont {Tabuteau},\ and\ \citenamefont
  {Pomeau}}]{mora2011surface}%
  \BibitemOpen
  \bibfield  {author} {\bibinfo {author} {\bibfnamefont {S.}~\bibnamefont
  {Mora}}, \bibinfo {author} {\bibfnamefont {M.}~\bibnamefont {Abkarian}},
  \bibinfo {author} {\bibfnamefont {H.}~\bibnamefont {Tabuteau}}, \ and\
  \bibinfo {author} {\bibfnamefont {Y.}~\bibnamefont {Pomeau}},\ }\href@noop {}
  {\bibfield  {journal} {\bibinfo  {journal} {Soft Matter}\ }\textbf {\bibinfo
  {volume} {7}},\ \bibinfo {pages} {10612} (\bibinfo {year}
  {2011})}\BibitemShut {NoStop}%
\bibitem [{\citenamefont {Gent}(1990)}]{gent1990cavitation}%
  \BibitemOpen
  \bibfield  {author} {\bibinfo {author} {\bibfnamefont {A.}~\bibnamefont
  {Gent}},\ }\href@noop {} {\bibfield  {journal} {\bibinfo  {journal} {Rubber
  Chemistry and Technology}\ }\textbf {\bibinfo {volume} {63}},\ \bibinfo
  {pages} {49} (\bibinfo {year} {1990})}\BibitemShut {NoStop}%
\bibitem [{\citenamefont {Shull}\ \emph {et~al.}(2000)\citenamefont {Shull},
  \citenamefont {Flanigan},\ and\ \citenamefont {Crosby}}]{shull2000fingering}%
  \BibitemOpen
  \bibfield  {author} {\bibinfo {author} {\bibfnamefont {K.}~\bibnamefont
  {Shull}}, \bibinfo {author} {\bibfnamefont {C.}~\bibnamefont {Flanigan}}, \
  and\ \bibinfo {author} {\bibfnamefont {A.}~\bibnamefont {Crosby}},\
  }\href@noop {} {\bibfield  {journal} {\bibinfo  {journal} {Phys. Rev. Lett.}\
  }\textbf {\bibinfo {volume} {84}},\ \bibinfo {pages} {3057} (\bibinfo {year}
  {2000})}\BibitemShut {NoStop}%
\bibitem [{\citenamefont {Saintyves}\ \emph {et~al.}(2013)\citenamefont
  {Saintyves}, \citenamefont {Dauchot},\ and\ \citenamefont
  {Bouchaud}}]{saintyves2013bulk}%
  \BibitemOpen
  \bibfield  {author} {\bibinfo {author} {\bibfnamefont {B.}~\bibnamefont
  {Saintyves}}, \bibinfo {author} {\bibfnamefont {O.}~\bibnamefont {Dauchot}},
  \ and\ \bibinfo {author} {\bibfnamefont {E.}~\bibnamefont {Bouchaud}},\
  }\href@noop {} {\bibfield  {journal} {\bibinfo  {journal} {Physical review
  letters}\ }\textbf {\bibinfo {volume} {111}},\ \bibinfo {pages} {047801}
  (\bibinfo {year} {2013})}\BibitemShut {NoStop}%
\bibitem [{\citenamefont {Biggins}\ \emph {et~al.}(2013)\citenamefont
  {Biggins}, \citenamefont {Saintyves}, \citenamefont {Wei}, \citenamefont
  {Bouchaud},\ and\ \citenamefont {Mahadevan}}]{bigginsfingering}%
  \BibitemOpen
  \bibfield  {author} {\bibinfo {author} {\bibfnamefont {J.}~\bibnamefont
  {Biggins}}, \bibinfo {author} {\bibfnamefont {B.}~\bibnamefont {Saintyves}},
  \bibinfo {author} {\bibfnamefont {A.}~\bibnamefont {Wei}}, \bibinfo {author}
  {\bibfnamefont {E.}~\bibnamefont {Bouchaud}}, \ and\ \bibinfo {author}
  {\bibfnamefont {L.}~\bibnamefont {Mahadevan}},\ }\href@noop {} {\bibfield
  {journal} {\bibinfo  {journal} {Proc.\ Nat.\ Acc.\ Sci.}\ } (\bibinfo {year}
  {2013})}\BibitemShut {NoStop}%
\bibitem [{\citenamefont {Biggins}\ \emph {et~al.}(2015)\citenamefont
  {Biggins}, \citenamefont {Wei},\ and\ \citenamefont
  {Mahadevan}}]{biggins2015fluid}%
  \BibitemOpen
  \bibfield  {author} {\bibinfo {author} {\bibfnamefont {J.~S.}\ \bibnamefont
  {Biggins}}, \bibinfo {author} {\bibfnamefont {Z.}~\bibnamefont {Wei}}, \ and\
  \bibinfo {author} {\bibfnamefont {L.}~\bibnamefont {Mahadevan}},\ }\href@noop
  {} {\bibfield  {journal} {\bibinfo  {journal} {EPL (Europhysics Letters)}\
  }\textbf {\bibinfo {volume} {110}},\ \bibinfo {pages} {34001} (\bibinfo
  {year} {2015})}\BibitemShut {NoStop}%
\bibitem [{\citenamefont {Mora}\ \emph {et~al.}(2014)\citenamefont {Mora},
  \citenamefont {Phou}, \citenamefont {Fromental},\ and\ \citenamefont
  {Pomeau}}]{mora2014gravity}%
  \BibitemOpen
  \bibfield  {author} {\bibinfo {author} {\bibfnamefont {S.}~\bibnamefont
  {Mora}}, \bibinfo {author} {\bibfnamefont {T.}~\bibnamefont {Phou}}, \bibinfo
  {author} {\bibfnamefont {J.-M.}\ \bibnamefont {Fromental}}, \ and\ \bibinfo
  {author} {\bibfnamefont {Y.}~\bibnamefont {Pomeau}},\ }\href@noop {}
  {\bibfield  {journal} {\bibinfo  {journal} {Physical review letters}\
  }\textbf {\bibinfo {volume} {113}},\ \bibinfo {pages} {178301} (\bibinfo
  {year} {2014})}\BibitemShut {NoStop}%
\bibitem [{\citenamefont {Liang}\ and\ \citenamefont
  {Cai}(2015)}]{liang2015gravity}%
  \BibitemOpen
  \bibfield  {author} {\bibinfo {author} {\bibfnamefont {X.}~\bibnamefont
  {Liang}}\ and\ \bibinfo {author} {\bibfnamefont {S.}~\bibnamefont {Cai}},\
  }\href@noop {} {\bibfield  {journal} {\bibinfo  {journal} {Applied Physics
  Letters}\ }\textbf {\bibinfo {volume} {106}},\ \bibinfo {pages} {041907}
  (\bibinfo {year} {2015})}\BibitemShut {NoStop}%
\bibitem [{\citenamefont {Hazel}\ and\ \citenamefont
  {Heil}(2005)}]{hazel2005surface}%
  \BibitemOpen
  \bibfield  {author} {\bibinfo {author} {\bibfnamefont {A.~L.}\ \bibnamefont
  {Hazel}}\ and\ \bibinfo {author} {\bibfnamefont {M.}~\bibnamefont {Heil}},\
  }\href@noop {} {\bibfield  {journal} {\bibinfo  {journal} {Proceedings of the
  Royal Society A: Mathematical, Physical and Engineering Science}\ }\textbf
  {\bibinfo {volume} {461}},\ \bibinfo {pages} {1847} (\bibinfo {year}
  {2005})}\BibitemShut {NoStop}%
\bibitem [{\citenamefont {Avery}\ and\ \citenamefont
  {Mead}(1959)}]{avery1959surface}%
  \BibitemOpen
  \bibfield  {author} {\bibinfo {author} {\bibfnamefont {M.~E.}\ \bibnamefont
  {Avery}}\ and\ \bibinfo {author} {\bibfnamefont {J.}~\bibnamefont {Mead}},\
  }\href@noop {} {\bibfield  {journal} {\bibinfo  {journal} {AMA journal of
  diseases of children}\ }\textbf {\bibinfo {volume} {97}},\ \bibinfo {pages}
  {517} (\bibinfo {year} {1959})}\BibitemShut {NoStop}%
\bibitem [{\citenamefont {Shyer}\ \emph {et~al.}(2013)\citenamefont {Shyer},
  \citenamefont {Tallinen}, \citenamefont {Nerurkar}, \citenamefont {Wei},
  \citenamefont {Gil}, \citenamefont {Kaplan}, \citenamefont {Tabin},\ and\
  \citenamefont {Mahadevan}}]{shyer2013villification}%
  \BibitemOpen
  \bibfield  {author} {\bibinfo {author} {\bibfnamefont {A.~E.}\ \bibnamefont
  {Shyer}}, \bibinfo {author} {\bibfnamefont {T.}~\bibnamefont {Tallinen}},
  \bibinfo {author} {\bibfnamefont {N.~L.}\ \bibnamefont {Nerurkar}}, \bibinfo
  {author} {\bibfnamefont {Z.}~\bibnamefont {Wei}}, \bibinfo {author}
  {\bibfnamefont {E.~S.}\ \bibnamefont {Gil}}, \bibinfo {author} {\bibfnamefont
  {D.~L.}\ \bibnamefont {Kaplan}}, \bibinfo {author} {\bibfnamefont {C.~J.}\
  \bibnamefont {Tabin}}, \ and\ \bibinfo {author} {\bibfnamefont
  {L.}~\bibnamefont {Mahadevan}},\ }\href@noop {} {\bibfield  {journal}
  {\bibinfo  {journal} {Science}\ }\textbf {\bibinfo {volume} {342}},\ \bibinfo
  {pages} {212} (\bibinfo {year} {2013})}\BibitemShut {NoStop}%
\bibitem [{\citenamefont {Hannezo}\ \emph {et~al.}(2011)\citenamefont
  {Hannezo}, \citenamefont {Prost},\ and\ \citenamefont
  {Joanny}}]{hannezo2011instabilities}%
  \BibitemOpen
  \bibfield  {author} {\bibinfo {author} {\bibfnamefont {E.}~\bibnamefont
  {Hannezo}}, \bibinfo {author} {\bibfnamefont {J.}~\bibnamefont {Prost}}, \
  and\ \bibinfo {author} {\bibfnamefont {J.-F.}\ \bibnamefont {Joanny}},\
  }\href@noop {} {\bibfield  {journal} {\bibinfo  {journal} {Physical Review
  Letters}\ }\textbf {\bibinfo {volume} {107}},\ \bibinfo {pages} {078104}
  (\bibinfo {year} {2011})}\BibitemShut {NoStop}%
\bibitem [{\citenamefont {Toro}\ and\ \citenamefont
  {Burnod}(2005)}]{toro2005morphogenetic}%
  \BibitemOpen
  \bibfield  {author} {\bibinfo {author} {\bibfnamefont {R.}~\bibnamefont
  {Toro}}\ and\ \bibinfo {author} {\bibfnamefont {Y.}~\bibnamefont {Burnod}},\
  }\href@noop {} {\bibfield  {journal} {\bibinfo  {journal} {Cerebral Cortex}\
  }\textbf {\bibinfo {volume} {15}},\ \bibinfo {pages} {1900} (\bibinfo {year}
  {2005})}\BibitemShut {NoStop}%
\bibitem [{\citenamefont {Tallinen}\ \emph {et~al.}(2014)\citenamefont
  {Tallinen}, \citenamefont {Chung}, \citenamefont {Biggins},\ and\
  \citenamefont {Mahadevan}}]{tallinen2014gyrification}%
  \BibitemOpen
  \bibfield  {author} {\bibinfo {author} {\bibfnamefont {T.}~\bibnamefont
  {Tallinen}}, \bibinfo {author} {\bibfnamefont {J.~Y.}\ \bibnamefont {Chung}},
  \bibinfo {author} {\bibfnamefont {J.~S.}\ \bibnamefont {Biggins}}, \ and\
  \bibinfo {author} {\bibfnamefont {L.}~\bibnamefont {Mahadevan}},\ }\href@noop
  {} {\bibfield  {journal} {\bibinfo  {journal} {Proceedings of the National
  Academy of Sciences}\ }\textbf {\bibinfo {volume} {111}},\ \bibinfo {pages}
  {12667} (\bibinfo {year} {2014})}\BibitemShut {NoStop}%
\bibitem [{\citenamefont {Bard}\ and\ \citenamefont {Ross}(1982)}]{BardRoss2}%
  \BibitemOpen
  \bibfield  {author} {\bibinfo {author} {\bibfnamefont {J.}~\bibnamefont
  {Bard}}\ and\ \bibinfo {author} {\bibfnamefont {A.}~\bibnamefont {Ross}},\
  }\href@noop {} {\bibfield  {journal} {\bibinfo  {journal} {Dev. Bio.}\
  }\textbf {\bibinfo {volume} {92}},\ \bibinfo {pages} {87} (\bibinfo {year}
  {1982})}\BibitemShut {NoStop}%
\bibitem [{\citenamefont {Savin}\ \emph {et~al.}(2011)\citenamefont {Savin},
  \citenamefont {Kurpios}, \citenamefont {Shyer}, \citenamefont {Florescu},
  \citenamefont {Liang}, \citenamefont {Mahadevan},\ and\ \citenamefont
  {Tabin}}]{savin2011growth}%
  \BibitemOpen
  \bibfield  {author} {\bibinfo {author} {\bibfnamefont {T.}~\bibnamefont
  {Savin}}, \bibinfo {author} {\bibfnamefont {N.~A.}\ \bibnamefont {Kurpios}},
  \bibinfo {author} {\bibfnamefont {A.~E.}\ \bibnamefont {Shyer}}, \bibinfo
  {author} {\bibfnamefont {P.}~\bibnamefont {Florescu}}, \bibinfo {author}
  {\bibfnamefont {H.}~\bibnamefont {Liang}}, \bibinfo {author} {\bibfnamefont
  {L.}~\bibnamefont {Mahadevan}}, \ and\ \bibinfo {author} {\bibfnamefont
  {C.~J.}\ \bibnamefont {Tabin}},\ }\href@noop {} {\bibfield  {journal}
  {\bibinfo  {journal} {Nature}\ }\textbf {\bibinfo {volume} {476}},\ \bibinfo
  {pages} {57} (\bibinfo {year} {2011})}\BibitemShut {NoStop}%
\bibitem [{\citenamefont {Li}\ \emph {et~al.}(2011)\citenamefont {Li},
  \citenamefont {Cao}, \citenamefont {Feng},\ and\ \citenamefont
  {Gao}}]{li2011surface}%
  \BibitemOpen
  \bibfield  {author} {\bibinfo {author} {\bibfnamefont {B.}~\bibnamefont
  {Li}}, \bibinfo {author} {\bibfnamefont {Y.-P.}\ \bibnamefont {Cao}},
  \bibinfo {author} {\bibfnamefont {X.-Q.}\ \bibnamefont {Feng}}, \ and\
  \bibinfo {author} {\bibfnamefont {H.}~\bibnamefont {Gao}},\ }\href@noop {}
  {\bibfield  {journal} {\bibinfo  {journal} {Journal of the Mechanics and
  Physics of Solids}\ }\textbf {\bibinfo {volume} {59}},\ \bibinfo {pages}
  {758} (\bibinfo {year} {2011})}\BibitemShut {NoStop}%
\bibitem [{\citenamefont {Tallinen}\ \emph {et~al.}(2013)\citenamefont
  {Tallinen}, \citenamefont {Biggins},\ and\ \citenamefont
  {Mahadevan}}]{tallinen2013surface}%
  \BibitemOpen
  \bibfield  {author} {\bibinfo {author} {\bibfnamefont {T.}~\bibnamefont
  {Tallinen}}, \bibinfo {author} {\bibfnamefont {J.}~\bibnamefont {Biggins}}, \
  and\ \bibinfo {author} {\bibfnamefont {L.}~\bibnamefont {Mahadevan}},\
  }\href@noop {} {\bibfield  {journal} {\bibinfo  {journal} {Phys. Rev. Lett.}\
  }\textbf {\bibinfo {volume} {110}},\ \bibinfo {pages} {024302} (\bibinfo
  {year} {2013})}\BibitemShut {NoStop}%
\bibitem [{\citenamefont {Tallinen}\ and\ \citenamefont
  {Biggins}(2015)}]{tallinen2015mechanics}%
  \BibitemOpen
  \bibfield  {author} {\bibinfo {author} {\bibfnamefont {T.}~\bibnamefont
  {Tallinen}}\ and\ \bibinfo {author} {\bibfnamefont {J.~S.}\ \bibnamefont
  {Biggins}},\ }\href {\doibase 10.1103/PhysRevE.92.022720} {\bibfield
  {journal} {\bibinfo  {journal} {Phys. Rev. E}\ }\textbf {\bibinfo {volume}
  {92}},\ \bibinfo {pages} {022720} (\bibinfo {year} {2015})}\BibitemShut
  {NoStop}%
\bibitem [{\citenamefont {Hannezo}\ \emph {et~al.}(2012)\citenamefont
  {Hannezo}, \citenamefont {Prost},\ and\ \citenamefont
  {Joanny}}]{hannezo2012mechanical}%
  \BibitemOpen
  \bibfield  {author} {\bibinfo {author} {\bibfnamefont {E.}~\bibnamefont
  {Hannezo}}, \bibinfo {author} {\bibfnamefont {J.}~\bibnamefont {Prost}}, \
  and\ \bibinfo {author} {\bibfnamefont {J.-F.}\ \bibnamefont {Joanny}},\
  }\href@noop {} {\bibfield  {journal} {\bibinfo  {journal} {Phys. Rev. Lett.}\
  }\textbf {\bibinfo {volume} {109}},\ \bibinfo {pages} {018101} (\bibinfo
  {year} {2012})}\BibitemShut {NoStop}%
\bibitem [{\citenamefont {Dervaux}\ \emph {et~al.}(2011)\citenamefont
  {Dervaux}, \citenamefont {Couder}, \citenamefont {Guedeau-Boudeville},\ and\
  \citenamefont {Ben~Amar}}]{dervaux2011shape}%
  \BibitemOpen
  \bibfield  {author} {\bibinfo {author} {\bibfnamefont {J.}~\bibnamefont
  {Dervaux}}, \bibinfo {author} {\bibfnamefont {Y.}~\bibnamefont {Couder}},
  \bibinfo {author} {\bibfnamefont {M.-A.}\ \bibnamefont {Guedeau-Boudeville}},
  \ and\ \bibinfo {author} {\bibfnamefont {M.}~\bibnamefont {Ben~Amar}},\
  }\href@noop {} {\bibfield  {journal} {\bibinfo  {journal} {Physical review
  letters}\ }\textbf {\bibinfo {volume} {107}},\ \bibinfo {pages} {018103}
  (\bibinfo {year} {2011})}\BibitemShut {NoStop}%
\bibitem [{\citenamefont {Cerda}\ and\ \citenamefont
  {Mahadevan}(2003)}]{cerda2003geometry}%
  \BibitemOpen
  \bibfield  {author} {\bibinfo {author} {\bibfnamefont {E.}~\bibnamefont
  {Cerda}}\ and\ \bibinfo {author} {\bibfnamefont {L.}~\bibnamefont
  {Mahadevan}},\ }\href@noop {} {\bibfield  {journal} {\bibinfo  {journal}
  {Physical review letters}\ }\textbf {\bibinfo {volume} {90}},\ \bibinfo
  {pages} {074302} (\bibinfo {year} {2003})}\BibitemShut {NoStop}%
\bibitem [{\citenamefont {Dervaux}\ and\ \citenamefont
  {Amar}(2011)}]{dervaux2011buckling}%
  \BibitemOpen
  \bibfield  {author} {\bibinfo {author} {\bibfnamefont {J.}~\bibnamefont
  {Dervaux}}\ and\ \bibinfo {author} {\bibfnamefont {M.~B.}\ \bibnamefont
  {Amar}},\ }\href@noop {} {\bibfield  {journal} {\bibinfo  {journal} {Journal
  of the Mechanics and Physics of Solids}\ }\textbf {\bibinfo {volume} {59}},\
  \bibinfo {pages} {538} (\bibinfo {year} {2011})}\BibitemShut {NoStop}%
\bibitem [{\citenamefont {Cai}\ and\ \citenamefont
  {Fu}(1999)}]{cai1999imperfection}%
  \BibitemOpen
  \bibfield  {author} {\bibinfo {author} {\bibfnamefont {Z.}~\bibnamefont
  {Cai}}\ and\ \bibinfo {author} {\bibfnamefont {Y.}~\bibnamefont {Fu}},\ }in\
  \href@noop {} {\emph {\bibinfo {booktitle} {Proceedings of the Royal Society
  of London A: Mathematical, Physical and Engineering Sciences}}},\ Vol.\
  \bibinfo {volume} {455}\ (\bibinfo {organization} {The Royal Society},\
  \bibinfo {year} {1999})\ pp.\ \bibinfo {pages} {3285--3309}\BibitemShut
  {NoStop}%
\bibitem [{\citenamefont {Cao}\ and\ \citenamefont
  {Hutchinson}(2012{\natexlab{a}})}]{cao2012wrinkling}%
  \BibitemOpen
  \bibfield  {author} {\bibinfo {author} {\bibfnamefont {Y.}~\bibnamefont
  {Cao}}\ and\ \bibinfo {author} {\bibfnamefont {J.~W.}\ \bibnamefont
  {Hutchinson}},\ }\href@noop {} {\bibfield  {journal} {\bibinfo  {journal}
  {Journal of applied mechanics}\ }\textbf {\bibinfo {volume} {79}},\ \bibinfo
  {pages} {031019} (\bibinfo {year} {2012}{\natexlab{a}})}\BibitemShut
  {NoStop}%
\bibitem [{\citenamefont {Cao}\ and\ \citenamefont
  {Hutchinson}(2012{\natexlab{b}})}]{cao2012wrinkles}%
  \BibitemOpen
  \bibfield  {author} {\bibinfo {author} {\bibfnamefont {Y.}~\bibnamefont
  {Cao}}\ and\ \bibinfo {author} {\bibfnamefont {J.~W.}\ \bibnamefont
  {Hutchinson}},\ }in\ \href@noop {} {\emph {\bibinfo {booktitle} {Proc. R.
  Soc. A}}},\ Vol.\ \bibinfo {volume} {468}\ (\bibinfo {organization} {The
  Royal Society},\ \bibinfo {year} {2012})\ pp.\ \bibinfo {pages}
  {94--115}\BibitemShut {NoStop}%
\bibitem [{\citenamefont {Allen}(1969)}]{allen1969}%
  \BibitemOpen
  \bibfield  {author} {\bibinfo {author} {\bibfnamefont {H.~G.}\ \bibnamefont
  {Allen}},\ }\href@noop {} {\emph {\bibinfo {title} {Analysis and Design of
  Structural Sandwich Panels: The Commonwealth and International Library:
  Structures and Solid Body Mechanics Division}}}\ (\bibinfo  {publisher}
  {Pergamon Press},\ \bibinfo {year} {1969})\BibitemShut {NoStop}%
\bibitem [{\citenamefont {Budday}\ \emph {et~al.}(2015)\citenamefont {Budday},
  \citenamefont {Kuhl},\ and\ \citenamefont {Hutchinson}}]{hutchinson_growth}%
  \BibitemOpen
  \bibfield  {author} {\bibinfo {author} {\bibfnamefont {S.}~\bibnamefont
  {Budday}}, \bibinfo {author} {\bibfnamefont {E.}~\bibnamefont {Kuhl}}, \ and\
  \bibinfo {author} {\bibfnamefont {J.~W.}\ \bibnamefont {Hutchinson}},\
  }\href@noop {} {\bibfield  {journal} {\bibinfo  {journal} {Philosophical
  Magazine}\ }\textbf {\bibinfo {volume} {95}},\ \bibinfo {pages} {3208}
  (\bibinfo {year} {2015})}\BibitemShut {NoStop}%
\bibitem [{\citenamefont {Biot}(1965)}]{biotbook}%
  \BibitemOpen
  \bibfield  {author} {\bibinfo {author} {\bibfnamefont {M.}~\bibnamefont
  {Biot}},\ }\href@noop {} {\emph {\bibinfo {title} {Mechanics of incremental
  deformations: theory of elasticity and viscoelasticity of initially stressed
  solids and fluids, including thermodynamic foundations and applications to
  finite strain}}}\ (\bibinfo  {publisher} {Wiley New York:},\ \bibinfo {year}
  {1965})\BibitemShut {NoStop}%
\bibitem [{\citenamefont {Hohlfeld}\ and\ \citenamefont
  {Mahadevan}(2011)}]{hohlfeld2011unfolding}%
  \BibitemOpen
  \bibfield  {author} {\bibinfo {author} {\bibfnamefont {E.}~\bibnamefont
  {Hohlfeld}}\ and\ \bibinfo {author} {\bibfnamefont {L.}~\bibnamefont
  {Mahadevan}},\ }\href@noop {} {\bibfield  {journal} {\bibinfo  {journal}
  {Phys. Rev. Lett.}\ }\textbf {\bibinfo {volume} {106}},\ \bibinfo {pages}
  {105702} (\bibinfo {year} {2011})}\BibitemShut {NoStop}%
\bibitem [{\citenamefont {Carroll}(2007)}]{carroll2007generating}%
  \BibitemOpen
  \bibfield  {author} {\bibinfo {author} {\bibfnamefont {M.}~\bibnamefont
  {Carroll}},\ }\href@noop {} {\bibfield  {journal} {\bibinfo  {journal}
  {Journal of Elasticity}\ }\textbf {\bibinfo {volume} {88}},\ \bibinfo {pages}
  {1} (\bibinfo {year} {2007})}\BibitemShut {NoStop}%
\bibitem [{\citenamefont {Ben~Amar}\ and\ \citenamefont
  {Ciarletta}(2010)}]{ben2010swelling}%
  \BibitemOpen
  \bibfield  {author} {\bibinfo {author} {\bibfnamefont {M.}~\bibnamefont
  {Ben~Amar}}\ and\ \bibinfo {author} {\bibfnamefont {P.}~\bibnamefont
  {Ciarletta}},\ }\href@noop {} {\bibfield  {journal} {\bibinfo  {journal}
  {Journal of the Mechanics and Physics of Solids}\ }\textbf {\bibinfo {volume}
  {58}},\ \bibinfo {pages} {935} (\bibinfo {year} {2010})}\BibitemShut
  {NoStop}%
\bibitem [{\citenamefont {Ciarletta}\ and\ \citenamefont
  {Ben~Amar}(2012{\natexlab{b}})}]{ciarletta2012papillary}%
  \BibitemOpen
  \bibfield  {author} {\bibinfo {author} {\bibfnamefont {P.}~\bibnamefont
  {Ciarletta}}\ and\ \bibinfo {author} {\bibfnamefont {M.}~\bibnamefont
  {Ben~Amar}},\ }\href@noop {} {\bibfield  {journal} {\bibinfo  {journal}
  {Mechanics Research Communications}\ }\textbf {\bibinfo {volume} {42}},\
  \bibinfo {pages} {68} (\bibinfo {year} {2012}{\natexlab{b}})}\BibitemShut
  {NoStop}%
\bibitem [{\citenamefont {Biggins}\ \emph
  {et~al.}(2014{\natexlab{a}})\citenamefont {Biggins}, \citenamefont {Wei},\
  and\ \citenamefont {Mahadevan}}]{biggins2014exactly}%
  \BibitemOpen
  \bibfield  {author} {\bibinfo {author} {\bibfnamefont {J.~S.}\ \bibnamefont
  {Biggins}}, \bibinfo {author} {\bibfnamefont {Z.}~\bibnamefont {Wei}}, \ and\
  \bibinfo {author} {\bibfnamefont {L.}~\bibnamefont {Mahadevan}},\ }\href@noop
  {} {\bibfield  {journal} {\bibinfo  {journal} {EPL (Europhysics Letters)}\
  }\textbf {\bibinfo {volume} {108}},\ \bibinfo {pages} {64001} (\bibinfo
  {year} {2014}{\natexlab{a}})}\BibitemShut {NoStop}%
\bibitem [{\citenamefont {Biggins}\ \emph
  {et~al.}(2014{\natexlab{b}})\citenamefont {Biggins}, \citenamefont {Wei},\
  and\ \citenamefont {Mahadevan}}]{biggins20142d}%
  \BibitemOpen
  \bibfield  {author} {\bibinfo {author} {\bibfnamefont {J.~S.}\ \bibnamefont
  {Biggins}}, \bibinfo {author} {\bibfnamefont {Z.}~\bibnamefont {Wei}}, \ and\
  \bibinfo {author} {\bibfnamefont {L.}~\bibnamefont {Mahadevan}},\ }\href@noop
  {} {\bibfield  {journal} {\bibinfo  {journal} {arXiv preprint
  arXiv:1407.1405}\ } (\bibinfo {year} {2014}{\natexlab{b}})}\BibitemShut
  {NoStop}%
\bibitem [{\citenamefont {Fleury}\ and\ \citenamefont
  {Gordon}(2012)}]{fleury2012coupling}%
  \BibitemOpen
  \bibfield  {author} {\bibinfo {author} {\bibfnamefont {V.}~\bibnamefont
  {Fleury}}\ and\ \bibinfo {author} {\bibfnamefont {R.}~\bibnamefont
  {Gordon}},\ }in\ \href@noop {} {\emph {\bibinfo {booktitle} {Origin (s) of
  Design in Nature}}}\ (\bibinfo  {publisher} {Springer},\ \bibinfo {year}
  {2012})\ pp.\ \bibinfo {pages} {385--428}\BibitemShut {NoStop}%
\end{thebibliography}
\end{document}